\documentclass[pra,reprint,nofootinbib,floatfix]{revtex4-2}
\usepackage[utf8]{inputenc}
\usepackage{color}
\usepackage{amsmath}
\usepackage{amssymb}
\usepackage{graphicx}
\usepackage[pdfusetitle,
 bookmarks=true,bookmarksnumbered=true,bookmarksopen=false,
 breaklinks=false,pdfborder={0 0 1},backref=false,colorlinks=true]
 {hyperref}

\makeatletter

\usepackage{amsfonts}
\usepackage{physics}
\usepackage{orcidlink}

\makeatother

\begin{document}
\title{Hamiltonian Conditions for Dark Modes in Multimode Bosonic Systems}
\author{Zuxin Wu$^{1}$}
\author{Shengshi Pang\orcidlink{0000-0002-6351-539X}$^{1,2}$}
\email{pangshsh@mail.sysu.edu.cn}

\affiliation{$^{1}$School of Physics, Sun Yat-sen University, Guangzhou, Guangdong
510275, China \\$^{2}$Hefei National Laboratory, University of Science
and Technology of China, Hefei 230088, China}
\begin{abstract}
Dark modes arise when destructive interference prevents selected bosonic
degrees of freedom from coupling to environmental channels. We formulate
a Hamiltonian criterion for identifying such modes in multimode bosonic
systems by separating two requirements: the candidate mode must be
invisible to the direct system--environment coupling, and its generated
operator space must remain invariant under the intrinsic system dynamics.
For linear environment coupling and quadratic system Hamiltonians,
the criterion is reduced to the the familiar null-space and invariant-subspace
conditions of passive linear dark-mode theory. We then extend the
analysis to nonlinear scenarios. For a two-photon conversion channel
coupled to an auxiliary environmental mode, interference among nonlinear
conversion pathways can reduce the environment coupling to a single
collective two-photon channel, leaving a complementary bosonic mode
decoupled from the environment. We show that preserving this mode
under nonlinear intrinsic dynamics generally requires more than conventional
Kerr-type quartic interactions: correlated four-boson conversion processes
are needed to cancel mixed nonlinear conversion between the dark and
environment-coupled collective modes. Finally, we show that the Bogoliubov
dark mode of a parametrically driven optomechanical satisfies the
same Hamiltonian criterion through an active canonical transformation.
These results provide a unified Hamiltonian framework for identifying
and engineering dark modes in linear, nonlinear, and driven bosonic
systems.
\end{abstract}
\maketitle

\section{Introduction}

\label{sec:introduction}

Uncontrolled coupling to environmental channels can dissipate excitations,
erase phase coherence, and degrade information encoded in a quantum
system. Protecting selected degrees of freedom from environmental
loss and decoherence is therefore a central problem in quantum optics
and open quantum systems. A standard route is destructive interference
among transition amplitudes into a lossy or radiative channel. In
driven atomic systems, such interference can produce a dark state:
a coherent superposition within the ground-state manifold whose optical
transition amplitude to an excited radiative state vanishes. This
mechanism underlies coherent population trapping \citep{grayCoherentTrappingAtomic1978,kellyDirectObservationCoherent2010,jamonneauCoherentPopulationTrapping2016,CoherentPopulationTrapping1996a}
and electromagnetically induced transparency \citep{bollerObservationElectromagneticallyInduced1991,ElectromagneticallyInducedTransparency1997,fleischhauerDarkStatePolaritonsElectromagnetically2000,fleischhauerQuantumMemoryPhotons2002,fleischhauerElectromagneticallyInducedTransparency2005,lukinColloquiumTrappingManipulating2003,zhaoGeneralDarkstateTheory2026a}.
A closely related interference principle also appears in collective
radiative systems, where different emission amplitudes can cancel
and give rise to subradiant Dicke-type states \citep{dickeCoherenceSpontaneousRadiation1954a,guerinSubradianceLargeCloud2016,bromleyCollectiveAtomicScattering2016}.

Multimode bosonic systems provide a natural setting for this interference-based
physics. In cavities, optomechanical devices, hybrid bosonic systems,
and related platforms, several bosonic modes can coexist within the
same device or couple to a shared environment. In such settings, the
environment may couple to a particular collective combination of the
physical modes rather than to each mode independently. The complementary
collective modes can then be absent from the direct system--environment
coupling.

This bright--dark mechanism has been explored in several bosonic
platforms. In optomechanical systems, dark modes have been used for
state transfer \citep{wangUsingInterferenceHigh2012,tianAdiabaticStateConversion2012},
optomechanically induced transparency \citep{dongOptomechanicalDarkMode2012,laiTunableOptomechanicallyInduced2020},
cooling \citep{huangMultimodeOptomechanicalCooling2022,liuGroundstateCoolingMultiple2022,caoOptomechanicalDarkModeBreakingCooling2025,wenSimultaneousGroundstateCooling2022},
and entanglement protocols \citep{huangThermalnoiseresistantOptomechanicalEntanglement2022,laiTripartiteOptomechanicalEntanglement2022}.
Recent dark-mode-engineering ideas have also been used to control
topological phonon blockade and its transfer in optomechanical networks
\citep{laiTopologicalPhononBlockade2025}. Related bright--dark structures
also appear in magnonic systems \citep{zhangMagnonDarkModes2015},
atom--cavity systems \citep{whiteCavityDarkMode2019,zhangCavityDarkMode2024},
and optomagnonic systems \citep{zhaoQuantumNetworksAssisted2023}.
More broadly, the same bright--dark viewpoint has also been used
to give a quantum description of classical optical interference \citep{villas-boasBrightDarkStates2025}.

From the theoretical side, related bright--dark decompositions, dark-mode
theorems, and decoherence-free constructions have been developed for
multimode and linear quantum systems \citep{benistyDarkModesSlow2009,yamamotoDecoherenceFreeLinearQuantum2014a,goughRealizationTheoryQuantum2015,panDarkModesQuantum2017,zhangKalmanDecompositionLinear2018,zhangLinearQuantumSystems2022,huangDarkModeTheoremsQuantum2023a}.
Across these settings, the compatibility between the environment coupling
and the system dynamics is usually encoded in the specific effective
Hamiltonian, rotating-frame description, resonance condition, control
protocol, or linear system matrix structure.

Here we formulate this compatibility directly as Hamiltonian conditions
for bosonic dark modes. For a general bosonic Hamiltonian, the absence
of direct system--environment coupling alone does not by itself specify
whether the corresponding mode remains dark under the full dynamics.
The intrinsic system Hamiltonian may still convert this mode into
modes that are directly coupled to the environment, so that the mode
couples to the environment indirectly. Thus a dark mode must satisfy
two requirements: it must be absent from the direct environment coupling,
and the intrinsic system Hamiltonian must not convert it into modes
coupled to the environment. Related requirements are formulated explicitly
in decoherence-free subspaces and noiseless subsystems \citep{zanardiNoiselessQuantumCodes1997a,zanardiErrorAvoidingQuantum1997,lidarDecoherenceFreeSubspacesQuantum1998,lidarConcatenatingDecoherenceFreeSubspaces1999,kempeTheoryDecoherencefreeFaulttolerant2001,lidarDecoherenceFreeSubspacesSubsystems2003,ticozziQuantumMarkovianSubsystems2008,ticozziAnalysisSynthesisAttractive2008,lidarReviewDecoherenceFree2014a},
where the protected sector must be compatible with both the noise
action and the internal dynamics.

In this work, we formulate these two requirements as Hamiltonian conditions
for multimode bosonic systems. We refer to them as environment invisibility
and dynamical invariance, respectively. Their exact operator forms
are given in Sec.~\ref{sec:hamiltonian_conditions}. The resulting
framework is not limited to linear models. It provides a direct method
for determining whether dark modes exist in a given Hamiltonian model
and, when they do, for deriving the corresponding mode operators.

As an application, we first examine the linear Hamiltonian model.
In this case, the general conditions simplify to the familiar matrix
structure in linear dark mode theory. We then apply the theory to
nonlinear system--environment couplings and nonlinear intrinsic interactions,
showing that the formation of dark modes requires compatible interference
in both parts of the Hamiltonian. Finally, we discuss the Bogoliubov
dark mode of a parametrically driven optomechanical interface as a
non-passive realization.

The rest of the paper is organized as follows. In Sec.~\ref{sec:dark_modes},
we introduce the intuitive physical picture of bosonic dark modes.
In Sec.~\ref{sec:hamiltonian_conditions}, we formulate the Hamiltonian
conditions for dark modes. In Sec.~\ref{sec:passive_quadratic},
we recover the dark-mode theory of linear models and discuss its minimal
two-mode realization. In Sec.~\ref{sec:nonlinear_dark_modes}, we
analyze dark modes in nonlinear models. In Sec.~\ref{sec:bogoliubov},
we study the optomechanical Bogoliubov dark mode as a non-passive
canonical realization of this Hamiltonian framework. Finally, Sec.~\ref{sec:conclusion}
summarizes the main results and discusses possible directions for
future work.

\section{Dark Modes in Multimode Bosonic Systems}

\label{sec:dark_modes}

We begin with the simplest two-mode example to build an intuitive
physical picture of dark modes in bosonic systems. The example shows
how destructive interference can make one collective mode absent from
the direct system--environment coupling, and why this absence alone
does not guarantee that the mode remains dark under the full dynamics.
It also introduces the terminology used below: we call such directly
decoupled modes environment-invisible modes, and the modes directly
coupled to the environment environment-coupled modes.

Consider two bosonic modes, $a_{1}$ and $a_{2}$, coupled to a common
environment through the rotating-wave exchange interaction 
\begin{equation}
H_{SE}=\sum_{k}\left(g_{k,1}a_{1}+g_{k,2}a_{2}\right)e^{\dagger}_{k}+\mathrm{H.c.},\label{eq:two_mode_intro_HSE}
\end{equation}
where $e_{k}$ is an environmental mode and $g_{k,j}$ is the coupling
amplitude between the system mode $a_{j}$ and the environmental mode
$e_{k}$. Suppose that all environmental modes couple to the same
relative combination of the two system modes, 
\begin{equation}
\frac{g_{k,1}}{g_{k,2}}=c,\label{eq:two_mode_intro_ratio}
\end{equation}
with $c$ independent of $k$. The environment then couples only to
a single collective combination of the bare modes, rather than to
two independent modes. The environment-coupled mode is 
\begin{equation}
B=\frac{ca_{1}+a_{2}}{\sqrt{1+|c|^{2}}}.\label{eq:two_mode_intro_B}
\end{equation}
The complementary environment-invisible mode is 
\begin{equation}
D=\frac{a_{1}-c^{\ast}a_{2}}{\sqrt{1+|c|^{2}}}.\label{eq:two_mode_intro_D}
\end{equation}
Together, $B$ and $D$ form two passive bosonic modes, 
\begin{equation}
{}[B,B^{\dagger}]=[D,D^{\dagger}]=1,\qquad[D,B^{\dagger}]=[D,B]=0.\label{eq:two_mode_intro_CCR}
\end{equation}
Thus $B$ and $D$ describe independent bosonic modes.

In terms of $B$ and $D$, the system--environment interaction becomes
\begin{equation}
H_{SE}=\sqrt{1+|c|^{2}}\left[B\left(\sum_{k}g_{k,2}e^{\dagger}_{k}\right)+B^{\dagger}\left(\sum_{k}g^{\ast}_{k,2}e_{k}\right)\right].\label{eq:two_mode_intro_HSE_BD}
\end{equation}
The direct coupling to the environment contains only $B$. The mode
$D$ is therefore environment-invisible. This is the standard interference
mechanism: the coupling amplitudes associated with $a_{1}$ and $a_{2}$
add coherently so that one collective mode is coupled to the bath,
while the complementary mode has zero direct coupling amplitude.

A coherent conversion term such as $B^{\dagger}D+D^{\dagger}B$, however,
can transfer an excitation initially stored in the environment-invisible
mode $D$ into the environment-coupled mode $B$. The mode $D$ then
becomes indirectly coupled to the environment through the intrinsic
system dynamics. Thus, a mode that is absent from the direct system--environment
interaction does not necessarily remain dark under the full dynamics.
We will derive the condition on the system Hamiltonian under which
$D$ becomes a dark mode in this two-mode model later.

The next section formulates this physical picture as Hamiltonian conditions
for canonical dark modes in bosonic systems. The central question
is how to decide, directly from the structure of the Hamiltonian,
whether a collection of canonical bosonic modes remains dark under
the full dynamics.

\section{Hamiltonian Conditions for Bosonic Dark Modes}

\label{sec:hamiltonian_conditions}

We now turn the physical picture of Sec.~\ref{sec:dark_modes} into
Hamiltonian conditions for bosonic dark modes. The central question
is how to decide, directly from the structure of the Hamiltonian,
whether the operator space generated by a collection of canonical
bosonic modes remains dark under the full Hamiltonian dynamics.

\subsection{Dark-mode operator space and Hamiltonian conditions}

From an observational perspective, a dark degree of freedom is one
whose observables evolve as if the environment were absent. To make
this statement precise, consider the Heisenberg evolution of a system
operator $O$ under the total Hamiltonian, 
\begin{equation}
O(t)=O+it[H_{{\rm tot}},O]+\frac{(it)^{2}}{2!}[H_{{\rm tot}},[H_{{\rm tot}},O]]+\cdots.\label{eq:Heisenberg_expansion}
\end{equation}
Thus, for the evolution of $O$ to be completely free of environmental
influence, none of the nested commutators should contain nontrivial
environmental operators.

Consider $n_{D}$ canonical bosonic modes $\{D_{k}\}^{n_{D}}_{k=1}$
satisfying 
\begin{equation}
{}[D_{k},D^{\dagger}_{l}]=\delta_{kl},\qquad[D_{k},D_{l}]=0,\qquad k,l=1,\dots,n_{D}.\label{eq:dark_mode_CCR}
\end{equation}
The physical observables of these modes are the Hermitian operators
constructed from $D_{k}$ and $D^{\dagger}_{k}$. For convenience,
we work with the complex operator space generated by the mode operators
and their adjoints, 
\begin{equation}
\mathcal{S}_{D}:=\left\{ F(D_{1},D^{\dagger}_{1},\ldots,D_{n_{D}},D^{\dagger}_{n_{D}})\right\} .\label{eq:dark_operator_space}
\end{equation}
Here $F$ denotes an arbitrary operator expression. We use the term
operator space to denote the collection of operators generated by
these mode operators. The Hermitian elements of $\mathcal{S}_{D}$
are the corresponding physical observables. Working with the full
operator space does not change the physical content. Any operator
$O\in\mathcal{S}_{D}$ can be decomposed as 
\begin{equation}
O=O_{1}+iO_{2},\qquad O_{1}=O^{\dagger}_{1},\qquad O_{2}=O^{\dagger}_{2},\label{eq:Hermitian_decomposition}
\end{equation}
with $O_{1},O_{2}\in\mathcal{S}_{D}$. Since any operator in $\mathcal{S}_{D}$
can be decomposed into Hermitian and anti-Hermitian parts, and since
the Heisenberg evolution is complex linear, requiring all Hermitian
observables to evolve without environmental operators is equivalent
to imposing the same requirement on the complex operator space $\mathcal{S}_{D}$.

The protected degree of freedom is therefore described not by a particular
choice of annihilation operators, but by the generated operator space
$\mathcal{S}_{D}$. The mode operators $D_{k}$ provide a canonical
set of generators for this space, but this choice need not be unique.
Different canonical annihilation and creation operators may generate
the same $\mathcal{S}_{D}$ and therefore describe the same dark degrees
of freedom. This generator freedom is discussed further in Appendix~\ref{app:generator_freedom}.

We now formulate the Hamiltonian condition for these modes to define
a dark mode. For these modes to remain dark modes under the full dynamics,
the operators generated by them must evolve within the same generated
operator space. If the evolution generated by $H_{{\rm tot}}$ takes
an operator in $\mathcal{S}_{D}$ outside $\mathcal{S}_{D}$, then
the dark degrees of freedom are no longer described by these modes
alone, even if the resulting operator contains no environmental operator.
We therefore formulate the dark-mode condition as 
\begin{align}
{}[H_{{\rm tot}},O_{D}]\in\mathcal{S}_{D},\qquad\forall O_{D}\in\mathcal{S}_{D}.\label{eq:full_dark_space_closure}
\end{align}
If this condition holds, then all higher nested commutators with $H_{{\rm tot}}$
also belong to $\mathcal{S}_{D}$. Hence they contain no nontrivial
environmental operators. Thus the operators of these modes evolve
within the same operator space and contain no nontrivial environmental
operators.

To express Eq.~\eqref{eq:full_dark_space_closure} in terms of the
Hamiltonian components, write 
\begin{equation}
H_{{\rm tot}}=H_{S}+H_{E}+H_{SE},\label{eq:Htot}
\end{equation}
where $H_{S}$ is the system Hamiltonian, $H_{E}$ describes the environment,
and $H_{SE}$ is the direct system--environment coupling. Since $O_{D}$
is a system operator, it commutes with $H_{E}$. Therefore 
\begin{equation}
{}[H_{{\rm tot}},O_{D}]=[H_{S},O_{D}]+[H_{SE},O_{D}].\label{eq:commutator_split}
\end{equation}
The first term is a system operator. The second term contains the
environmental operator structures appearing in the direct coupling.
Assuming that all purely system terms have been included in $H_{S}$,
these environmental structures cannot be cancelled by $[H_{S},O_{D}]$.
Therefore Eq.~\eqref{eq:full_dark_space_closure} gives two requirements:
\begin{equation}
{}[H_{SE},O_{D}]=0,\qquad\forall O_{D}\in\mathcal{S}_{D},\label{eq:environmental_invisibility_full}
\end{equation}
and 
\begin{equation}
{}[H_{S},O_{D}]\in\mathcal{S}_{D},\qquad\forall O_{D}\in\mathcal{S}_{D}.\label{eq:dynamical_closure_full}
\end{equation}
We call Eq.~\eqref{eq:environmental_invisibility_full} environment
invisibility. It states that the direct system--environment coupling
does not probe the dark-mode degrees of freedom. We call Eq.~\eqref{eq:dynamical_closure_full}
dynamical invariance. It states that the intrinsic system Hamiltonian
may generate nontrivial dynamics within the dark modes, but it must
not convert the operators of $\mathcal{S}_{D}$ into operators outside
$\mathcal{S}_{D}$.

\subsection{Environment invisibility}

We first discuss the direct environment-invisibility condition. Since
$\mathcal{S}_{D}$ is generated by the mode operators $D_{k}$ and
$D^{\dagger}_{k}$, Eq.~\eqref{eq:environmental_invisibility_full}
can be checked on the generators. Indeed, if 
\begin{equation}
{}[H_{SE},D_{k}]=[H_{SE},D^{\dagger}_{k}]=0,\qquad k=1,\dots,n_{D},\label{eq:environmental_invisibility_generators}
\end{equation}
then, by the Leibniz rule for commutators, every operator constructed
from these generators also commutes with $H_{SE}$. Conversely, if
Eq.~\eqref{eq:environmental_invisibility_full} holds for all $O_{D}\in\mathcal{S}_{D}$,
then it holds in particular for the generators. Therefore Eq.~\eqref{eq:environmental_invisibility_generators}
is equivalent to Eq.~\eqref{eq:environmental_invisibility_full}.

To make the meaning of this condition more explicit, complete the
$D$ modes to a full bosonic mode basis. The corresponding change
of modes can be represented by a system unitary $\mathcal{U}$, so
that 
\begin{equation}
D_{k}=\mathcal{U}a_{k}\mathcal{U}^{\dagger},\qquad k=1,\dots,n_{D},\label{eq:general_canonical_dark_mode}
\end{equation}
and the remaining modes are defined by 
\begin{equation}
B_{\alpha}=\mathcal{U}a_{n_{D}+\alpha}\mathcal{U}^{\dagger},\qquad\alpha=1,\dots,n_{B},\qquad n_{B}=n-n_{D}.\label{eq:general_canonical_bright_mode}
\end{equation}
The $B_{\alpha}$ are the complementary modes. The full set $\{D_{k},B_{\alpha}\}$
satisfies 
\begin{equation}
\begin{aligned}{}[D_{k},D^{\dagger}_{l}] & =\delta_{kl}, & {}[D_{k},D_{l}] & =0,\\{}
[B_{\alpha},B^{\dagger}_{\beta}] & =\delta_{\alpha\beta}, & {}[B_{\alpha},B_{\beta}] & =0,\\{}
[D_{k},B^{\dagger}_{\alpha}] & =0, & {}[D_{k},B_{\alpha}] & =0.
\end{aligned}
\label{eq:dark_CCR}
\end{equation}

In this notation, environment invisibility reads 
\begin{equation}
\begin{aligned}{}[H_{SE},\mathcal{U}a_{k}\mathcal{U}^{\dagger}] & =0,\\{}
[H_{SE},\mathcal{U}a^{\dagger}_{k}\mathcal{U}^{\dagger}] & =0,
\end{aligned}
\qquad k=1,\dots,n_{D}.\label{eq:general_U_dark_condition}
\end{equation}
Equivalently, after transforming the interaction Hamiltonian to the
reference modes, one obtains 
\begin{equation}
\begin{aligned}{}[\mathcal{U}^{\dagger}H_{SE}\mathcal{U},a_{k}] & =0,\\{}
[\mathcal{U}^{\dagger}H_{SE}\mathcal{U},a^{\dagger}_{k}] & =0,
\end{aligned}
\qquad k=1,\dots,n_{D}.\label{eq:general_U_dark_condition_transformed}
\end{equation}
Thus the search for environment-invisible modes can be formulated
either as an operator equation for the transformed mode operators,
as in Eq.~\eqref{eq:general_U_dark_condition}, or as a condition
on the transformed interaction Hamiltonian, as in Eq.~\eqref{eq:general_U_dark_condition_transformed}.

Environment invisibility alone, however, does not guarantee darkness
under the full dynamics. The intrinsic system Hamiltonian may still
convert operators in $\mathcal{S}_{D}$ into operators involving the
complementary modes $B_{\alpha}$. The operators in $\mathcal{S}_{D}$
would then become affected by the environment indirectly, and the
modes would not remain dark. To avoid this coherent conversion, we
should consider the second condition.

\subsection{Dynamical invariance}

The second condition is dynamical invariance, 
\begin{equation}
{}[H_{S},O_{D}]\in\mathcal{S}_{D},\qquad\forall O_{D}\in\mathcal{S}_{D}.\label{eq:dynamical_closure_full_repeat}
\end{equation}
It ensures that the intrinsic system dynamics preserves the dark-mode
operator space. Again, it is enough to check the generators. If 
\begin{equation}
{}[H_{S},D_{k}]\in\mathcal{S}_{D},\qquad[H_{S},D^{\dagger}_{k}]\in\mathcal{S}_{D},\qquad k=1,\dots,n_{D},\label{eq:dynamical_closure_generators}
\end{equation}
then the Leibniz rule implies $[H_{S},O_{D}]\in\mathcal{S}_{D}$ for
every $O_{D}\in\mathcal{S}_{D}$. Conversely, Eq.~\eqref{eq:dynamical_closure_full_repeat}
immediately implies Eq.~\eqref{eq:dynamical_closure_generators},
because the generators themselves belong to $\mathcal{S}_{D}$.

Dynamical invariance also has a simple Hamiltonian interpretation
in the $\{D_{k},B_{\alpha}\}$ mode frame. The system Hamiltonian
may generate nontrivial dynamics within the $D$ modes and within
the complementary $B$ modes separately, but it must not contain processes
that mix the $D$ modes with the $B$ modes. Equivalently, it can
be written as 
\begin{equation}
\begin{split}H_{S}={} & H_{D}\bigl(D_{1},D^{\dagger}_{1},\ldots,D_{n_{D}},D^{\dagger}_{n_{D}}\bigr)\\
 & +H_{B}\bigl(B_{1},B^{\dagger}_{1},\ldots,B_{n_{B}},B^{\dagger}_{n_{B}}\bigr).
\end{split}
\label{eq:separated_Hamiltonian_form}
\end{equation}
Here $H_{D}$ contains only the $D$-mode operators, while $H_{B}$
contains only the complementary $B$-mode operators. In this form,
operators in $\mathcal{S}_{D}$ may evolve nontrivially under $H_{D}$,
but the system Hamiltonian cannot convert them into operators involving
the complementary modes. The implication from Eq.~\eqref{eq:separated_Hamiltonian_form}
to dynamical invariance is immediate; the converse statement is justified
in Appendix~\ref{app:closure}.

Equation~\eqref{eq:separated_Hamiltonian_form} therefore provides
an equivalent Hamiltonian test once the mode frame has been found.
In general, however, solving Eqs.~\eqref{eq:general_U_dark_condition}
or \eqref{eq:general_U_dark_condition_transformed} for an arbitrary
canonical transformation $\mathcal{U}$ is a difficult operator-level
problem. Concrete models are therefore most naturally treated by choosing
an ansatz adapted to the structure of the system--environment coupling.
Depending on the model, this ansatz may reduce to a passive collective-mode
transformation or may require a more general Bogoliubov-type transformation,
as illustrated in the examples below.

\section{Linear Model and Passive Dark Modes}

\label{sec:passive_quadratic}

The Hamiltonian conditions introduced in Sec.~\ref{sec:hamiltonian_conditions}
apply to general dark modes. It is nevertheless useful to first examine
the passive linear setting, where the dark modes are passive superpositions
of the bare bosonic modes. This setting directly connects with the
standard bright--dark construction in linear dark-mode theory and
provides a clear reference point for the nonlinear examples discussed
later.

We therefore consider a system with a quadratic system Hamiltonian
and a linear system--environment coupling. In this framework, the
two physical requirements---environment invisibility and dynamical
invariance---reduce to simple matrix conditions. Dark-mode constructions
and decoherence-free structures in quantum linear systems have been
studied extensively\citep{yamamotoDecoherenceFreeLinearQuantum2014a,goughRealizationTheoryQuantum2015,zhangKalmanDecompositionLinear2018,zhangLinearQuantumSystems2022}.
The purpose of this section is not to develop a new theory of linear
dark modes, but to show explicitly how this known structure follows
from the general Hamiltonian conditions formulated above.

\subsection{Passive linear reduction}

We first consider the system--bath interaction. For bosonic modes
weakly coupled to a bath, the leading exchange-type coupling after
the rotating-wave approximation is linear in the annihilation and
creation operators. We write 
\begin{equation}
H_{SE}=\sum_{\mu}\left(\sum^{n}_{j=1}g_{\mu j}a_{j}\right)e^{\dagger}_{\mu}+\mathrm{H.c.},\label{eq:linear_HSE}
\end{equation}
where $e_{\mu}$ denotes an environmental mode and $g_{\mu j}$ is
the coupling amplitude between the system mode $a_{j}$ and the environmental
mode $e_{\mu}$. These amplitudes define the coupling matrix 
\begin{equation}
G=\begin{pmatrix}g_{11} & g_{12} & \cdots & g_{1n}\\
g_{21} & g_{22} & \cdots & g_{2n}\\
\vdots & \vdots & \ddots & \vdots\\
g_{m1} & g_{m2} & \cdots & g_{mn}
\end{pmatrix}.\label{eq:coupling_matrix_G}
\end{equation}

We take a passive collective mode to have the form 
\begin{equation}
D_{k}=\sum^{n}_{j=1}v^{*}_{kj}a_{j},\qquad k=1,\dots,n_{D},\label{eq:passive_dark_generator}
\end{equation}
with coefficient vector 
\begin{equation}
v_{k}=\left(v_{k1},v_{k2},\ldots,v_{kn}\right)^{T}.\label{eq:passive_dark_vector}
\end{equation}
When the coefficient vectors are chosen to satisfy 
\begin{equation}
v^{\dagger}_{k}v_{l}=\delta_{kl},\label{eq:passive_vector_normalization}
\end{equation}
the corresponding operators obey 
\begin{equation}
{}[D_{k},D^{\dagger}_{l}]=\delta_{kl},\qquad[D_{k},D_{l}]=0.\label{eq:passive_dark_CCR}
\end{equation}
Thus the vectors $v_{k}$ define independent passive bosonic modes.

Applying the direct-decoupling condition in Eq.~\eqref{eq:environmental_invisibility_generators}
to Eq.~\eqref{eq:passive_dark_generator} gives 
\begin{equation}
{}[H_{SE},D_{k}]=-\sum_{\mu}\left(\sum^{n}_{j=1}g_{\mu j}v_{kj}\right)^{*}e_{\mu}.\label{eq:linear_HSE_commutator}
\end{equation}
Hence 
\begin{equation}
{}[H_{SE},D_{k}]=[H_{SE},D^{\dagger}_{k}]=0\quad\Longleftrightarrow\quad Gv_{k}=0.\label{eq:linear_dark_vector_condition}
\end{equation}
The meaning is straightforward: a passive collective mode is absent
from the environmental channel when its coupling amplitude vanishes
for every environmental mode. Thus, for linear bath coupling, the
environment-invisible passive modes are the solutions of $Gv=0$.
If 
\begin{equation}
\operatorname{rank}(G)=r,\label{eq:rank_G}
\end{equation}
then there are $n-r$ independent passive combinations with this property.

We next take the system Hamiltonian to be quadratic, 
\begin{equation}
H_{S}=\sum^{n}_{i,j=1}\Omega_{ij}a^{\dagger}_{i}a_{j},\qquad\Omega=\Omega^{\dagger}.\label{eq:quadratic_HS}
\end{equation}
Here $\Omega$ contains the bare mode frequencies and coherent mode
couplings. For a passive mode $D_{k}$ of the form in Eq.~\eqref{eq:passive_dark_generator},
one finds 
\begin{equation}
{}[H_{S},D_{k}]=-\sum^{n}_{j=1}\left[(\Omega v_{k})_{j}\right]^{*}a_{j}.\label{eq:quadratic_action_on_Dk}
\end{equation}
Thus, the commutator of the quadratic Hamiltonian with a passive collective
annihilation operator is again a passive collective annihilation operator.
In terms of the coefficient vector, the action of $H_{S}$ is represented
by multiplication with the quadratic Hamiltonian matrix $\Omega$.

Let 
\begin{equation}
\{v_{1},v_{2},\ldots,v_{n-r}\}\label{eq:bath_invisible_basis}
\end{equation}
be a complete set of coefficient vectors satisfying $Gv_{k}=0$, and
define 
\begin{equation}
V_{D}:=\operatorname{span}\{v_{1},v_{2},\ldots,v_{n-r}\}.\label{eq:dark_vector_subspace}
\end{equation}
For the corresponding passive modes to remain dark, the intrinsic
dynamics must preserve this environment-invisible collective-mode
subspace rather than rotate it into environment-coupled modes. By
Eq.~\eqref{eq:quadratic_action_on_Dk}, this is equivalent to requiring
\begin{equation}
\Omega v_{k}\in V_{D},\qquad k=1,\dots,n-r.\label{eq:quadratic_dark_preservation_subspace}
\end{equation}
In other words, the coefficient space selected by $Gv=0$ must be
invariant under the matrix $\Omega$.

The passive linear conditions can therefore be summarized as 
\begin{equation}
Gv_{k}=0,\qquad\Omega v_{k}\in V_{D}.\label{eq:linear_dark_summary}
\end{equation}
The first equation corresponds to environment invisibility, and the
second equation corresponds to dynamical invariance. This is the usual
matrix structure underlying passive dark modes and related decoherence-free
constructions in linear quantum systems \citep{yamamotoDecoherenceFreeLinearQuantum2014a,panDarkModesQuantum2017,huangDarkModeTheoremsQuantum2023a}.
The minimal two-mode realization below makes the physical meaning
of these two conditions explicit.

\subsection{Minimal two-mode realization and interference-based decoupling}

We now return to the two-mode bright--dark construction introduced
in Sec.~\ref{sec:dark_modes}. There, the common-bath coupling satisfies
the fixed-ratio condition in Eq.~\eqref{eq:two_mode_intro_ratio},
so that the environment couples only to the mode $B$ and the complementary
mode $D$ is defined in Eqs.~\eqref{eq:two_mode_intro_B} and \eqref{eq:two_mode_intro_D}.
As shown in Eq.~\eqref{eq:two_mode_intro_HSE_BD}, the direct coupling
contains only $B$, and therefore $D$ is absent from the system--environment
coupling.

The remaining question is whether a concrete system Hamiltonian prevents
the mode $D$ from being converted into the mode $B$. After choosing
a relative phase convention in which the coherent hopping amplitude
is real, the quadratic Hamiltonian can be written as 
\begin{equation}
H_{S}=\omega_{1}a^{\dagger}_{1}a_{1}+\omega_{2}a^{\dagger}_{2}a_{2}+\kappa\left(a^{\dagger}_{1}a_{2}+a^{\dagger}_{2}a_{1}\right),\label{eq:two_mode_HS}
\end{equation}
where $\omega_{1}$ and $\omega_{2}$ are the bare mode frequencies
and $\kappa$ is the coherent hopping amplitude. It is convenient
to introduce 
\begin{equation}
\omega_{0}=\frac{\omega_{1}+\omega_{2}}{2},\qquad\Delta=\frac{\omega_{1}-\omega_{2}}{2}.\label{eq:omega0_Delta}
\end{equation}
Expressed in terms of the $B$ and $D$ modes, the system Hamiltonian
becomes 
\begin{equation}
\begin{aligned}H_{S} & =\omega_{B}B^{\dagger}B+\omega_{D}D^{\dagger}D\\
 & \quad+\left[\frac{2c\Delta+\kappa(1-c^{2})}{1+|c|^{2}}B^{\dagger}D+\mathrm{H.c.}\right],
\end{aligned}
\label{eq:two_mode_HS_BD_compact}
\end{equation}
where 
\begin{equation}
\omega_{B}=\omega_{0}+\frac{(|c|^{2}-1)\Delta+\kappa(c+c^{*})}{1+|c|^{2}},\label{eq:omega_B_general}
\end{equation}
and 
\begin{equation}
\omega_{D}=\omega_{0}-\frac{(|c|^{2}-1)\Delta+\kappa(c+c^{*})}{1+|c|^{2}}.\label{eq:omega_D_general}
\end{equation}
The final term in Eq.~\eqref{eq:two_mode_HS_BD_compact} is the coherent
$B$--$D$ conversion generated by the intrinsic system Hamiltonian.
Therefore, the environment-invisible mode $D$ becomes dark if and
only if 
\begin{equation}
2c\Delta+\kappa(1-c^{2})=0.\label{eq:BIC_condition}
\end{equation}
In this case, $D$ is an independently evolving dark mode.

\begin{figure}[t]
\centering \includegraphics[width=0.95\linewidth]{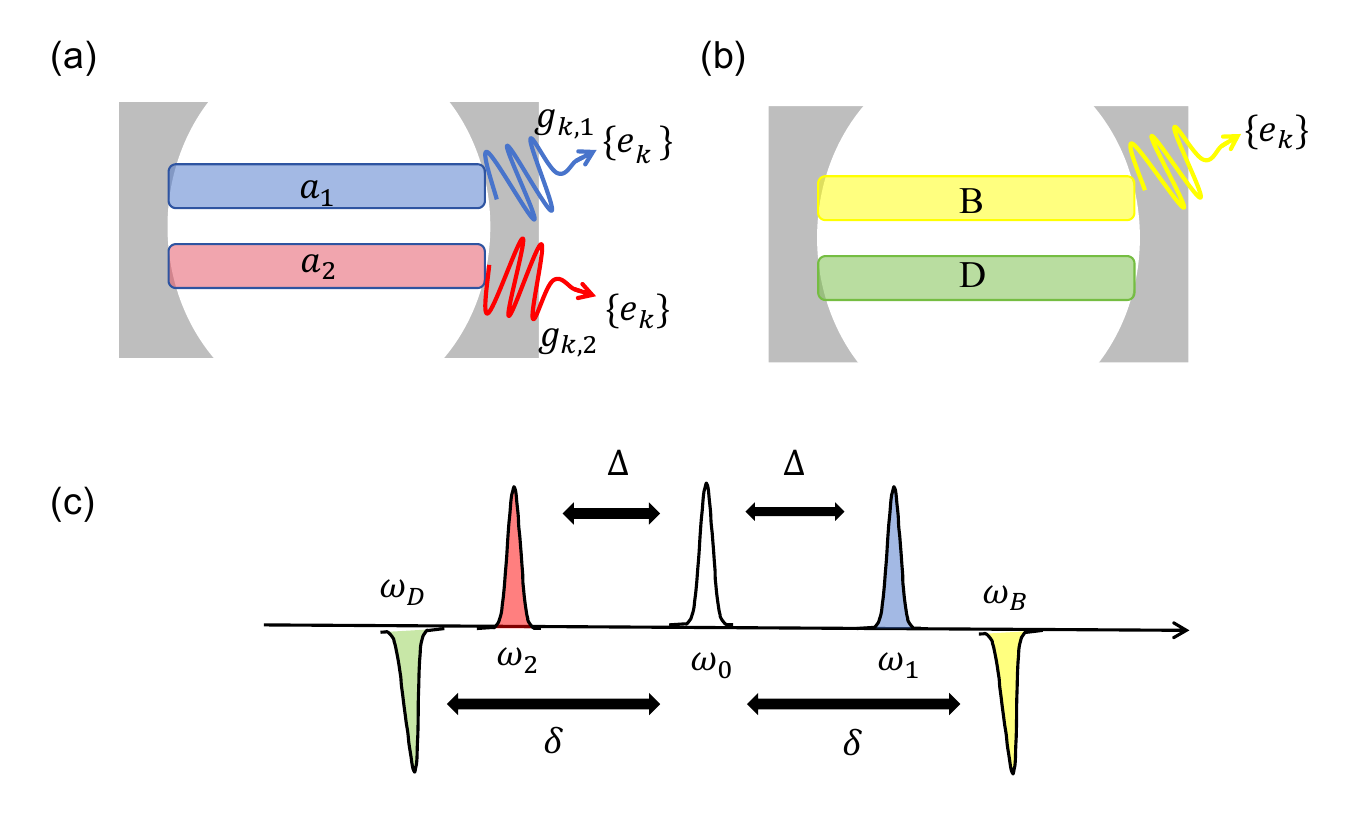} \caption{Passive two-mode $B$--$D$ structure and the associated dynamical-invariance
condition. (a) Two bosonic modes $a_{1}$ and $a_{2}$ couple to a
common set of environmental modes $\{e_{k}\}$ with amplitudes $g_{k,1}$
and $g_{k,2}$. When the ratio $g_{k,1}/g_{k,2}=c$ is independent
of $k$, the environment couples only to one collective combination
of the two modes. (b) In the $B$--$D$ basis introduced in Sec.~\ref{sec:dark_modes},
only the bright mode $B$ remains coupled directly to the environment,
while the mode $D$ is absent from the system--environment interaction.
(c) Schematic frequency structure. The bare-mode frequencies $\omega_{1}$
and $\omega_{2}$ are displaced from their mean $\omega_{0}$ by $\Delta$.
Once the intrinsic Hamiltonian also satisfies Eq.~\eqref{eq:BIC_condition},
the bright and dark modes become the normal modes of $H_{S}$. For
the ordering shown here, their frequencies are $\omega_{B}=\omega_{0}+\delta$
and $\omega_{D}=\omega_{0}-\delta$, with $\delta=\sqrt{\Delta^{2}+\kappa^{2}}$.}
\label{fig:linear}
\end{figure}

Equation~\eqref{eq:BIC_condition} has the same form as the destructive-interference
condition that appears in Friedrich--Wintgen-type bound states in
the continuum \citep{friedrichInterferingResonancesBound1985,zhenTopologicalNatureOptical2014,hsuBoundStatesContinuum2016,koshelevBoundStatesContinuum2023,wangOpticalBoundStates2024}.
In the present two-mode language, the condition states that the eigenmode
selected by the Hamiltonian must coincide with the environment-invisible
mode. This gives a direct connection between dark-mode formation and
interference-based decoupling in open optical systems.

When Eq.~\eqref{eq:BIC_condition} holds, the corresponding frequency
splitting is 
\begin{equation}
|\omega_{B}-\omega_{D}|=2\sqrt{\Delta^{2}+\kappa^{2}}.\label{eq:frequency_splitting}
\end{equation}
For the ordering depicted in Fig.~\ref{fig:linear}(c), we denote
the positive half-splitting by 
\begin{equation}
\delta=\sqrt{\Delta^{2}+\kappa^{2}},\label{eq:delta_linear}
\end{equation}
so that 
\begin{equation}
\omega_{B}=\omega_{0}+\delta,\qquad\omega_{D}=\omega_{0}-\delta.\label{eq:omega_BD_split}
\end{equation}
In the symmetric-coupling limit $c=1$, Eq.~\eqref{eq:BIC_condition}
reduces to the resonance condition 
\begin{equation}
\Delta=0,\label{eq:symmetric_resonance}
\end{equation}
with arbitrary $\kappa$.

The coupling structure of the passive two-mode model, its $B$--$D$
mode reduction, and the associated frequency splitting under the dynamical-invariance
condition are summarized schematically in Fig.~\ref{fig:linear}.

Because the $B$--$D$ transformation in this case is passive, the
corresponding modes can in principle be accessed by linear mode mixing,
such as beam-splitter-type transformations or their multimode generalizations.
This passive two-mode model will serve as the reference point for
the nonlinear examples developed below.

\section{Nonlinear Model and Nonlinear Interference}

\label{sec:nonlinear_dark_modes}

The passive quadratic benchmark of Sec.~\ref{sec:passive_quadratic}
illustrates the simplest application of the Hamiltonian conditions.
In that setting, the environment couples to one collective mode and
leaves a complementary environment-invisible mode, while the intrinsic
system dynamics preserves this mode and prevents it from being converted
into the environment-coupled mode. In nonlinear optical and circuit
systems, however, system--environment coupling can also be mediated
by conversion processes involving products of field operators\citep{leghtasConfiningStateLight2015b,miranowiczStatedependentPhotonBlockade2014,puriEngineeringQuantumStates2017}.
In such cases, the system--environment coupling need not be linear
in a single field amplitude, but may involve a composite amplitude,
such as a two-photon pair operator.

Nonlinear conversion processes lead to the same interference problem
in a different form\citep{yelinNonlinearOpticsDouble2003,niuGiantKerrNonlinearity2005,scheuerAllopticalSwitchingDark2010,huangDarkStateNonlinear2015}.
Several coherent conversion pathways can feed the same auxiliary environmental
mode. Their relative phases and amplitudes determine whether the auxiliary
mode couples to several independent two-photon conversion channels
or to a single collective two-photon channel. In the latter case,
a complementary bosonic mode can be identified that is absent from
the nonlinear process. The remaining question is whether the system
Hamiltonian satisfies dynamical invariance, namely, whether it avoids
converting the environment-invisible mode into the environment-coupled
mode.

In this section, we first use a two-photon conversion channel to show
how a nonlinear process can select an environment-invisible bosonic
mode. We then examine which quartic system Hamiltonians keep the corresponding
environment-invisible mode dynamically separated from the environment-coupled
mode. The result is a nonlinear interference condition: both the system--environment
conversion process and the intrinsic nonlinear interactions must be
arranged so that mixed conversion amplitudes cancel.

\subsection{Environment-invisible modes selected by a nonlinear system--environment
coupling}

Consider two bosonic system modes, $a_{1}$ and $a_{2}$, coupled
to an auxiliary mode $e$ representing an environmental channel through
a two-photon conversion process, 
\begin{equation}
H^{(2)}_{SE}=\left(\chi_{1}a^{2}_{1}+\chi_{12}a_{1}a_{2}+\chi_{2}a^{2}_{2}\right)e^{\dagger}+\mathrm{H.c.}\label{eq:nonlinear_HSE_two_photon}
\end{equation}
The coefficients $\chi_{1}$, $\chi_{12}$, and $\chi_{2}$ are the
conversion amplitudes associated with three distinct pathways by which
two system excitations are converted into one excitation of the auxiliary
environmental mode. Their relative phases and magnitudes determine
whether the auxiliary mode distinguishes these microscopic pathways
or couples only to one collective two-photon conversion channel.

To find a bosonic mode that is absent from this nonlinear channel,
we allow a general two-mode Bogoliubov form 
\begin{equation}
D=u_{1}a_{1}+u_{2}a_{2}+w_{1}a^{\dagger}_{1}+w_{2}a^{\dagger}_{2}.\label{eq:nonlinear_HSE_Bogoliubov_ansatz}
\end{equation}
The condition that this mode does not enter the direct coupling to
the auxiliary environmental mode is 
\begin{equation}
{}[H^{(2)}_{SE},D]=[H^{(2)}_{SE},D^{\dagger}]=0.\label{eq:nonlinear_HSE_dark_condition_general}
\end{equation}
Substituting Eq.~\eqref{eq:nonlinear_HSE_Bogoliubov_ansatz} gives
\begin{equation}
\begin{pmatrix}2\chi_{1} & \chi_{12}\\
\chi_{12} & 2\chi_{2}
\end{pmatrix}\begin{pmatrix}w_{1}\\
w_{2}
\end{pmatrix}=0,\qquad\begin{pmatrix}2\chi^{*}_{1} & \chi^{*}_{12}\\
\chi^{*}_{12} & 2\chi^{*}_{2}
\end{pmatrix}\begin{pmatrix}u_{1}\\
u_{2}
\end{pmatrix}=0.\label{eq:nonlinear_HSE_matrix_conditions}
\end{equation}
Thus, a nontrivial environment-invisible solution exists only when
the two-photon conversion matrix has rank one. Equivalently, the three
pair-conversion amplitudes must be matched so as to form the square
of a single collective field. This requires 
\begin{equation}
4\chi_{1}\chi_{2}-\chi^{2}_{12}=0.\label{eq:nonlinear_HSE_rank_one_condition}
\end{equation}
For $\chi_{2}\neq0$, this can be written as 
\begin{equation}
\chi_{1}=c^{2}\chi_{2},\qquad\chi_{12}=2c\chi_{2},\label{eq:nonlinear_HSE_conditions}
\end{equation}
with complex $c$.

Under Eq.~\eqref{eq:nonlinear_HSE_conditions}, the auxiliary environmental
mode no longer distinguishes the three microscopic pair-conversion
pathways. Their amplitudes interfere into one collective two-photon
channel. A convenient passive representative of the environment-invisible
mode is 
\begin{equation}
d=\frac{a_{1}-c^{*}a_{2}}{\sqrt{1+|c|^{2}}}.\label{eq:nonlinear_HSE_passive_dark_representative}
\end{equation}
As discussed in Sec.~\ref{sec:hamiltonian_conditions}, the same
generated operator space need not correspond to a unique choice of
annihilation and creation operators. In the present single-mode case,
the environment-invisible mode may more generally be represented as
\begin{equation}
D=\mu d+\nu d^{\dagger},\qquad|\mu|^{2}-|\nu|^{2}=1.\label{eq:nonlinear_HSE_dark_reparametrization}
\end{equation}
Thus the nonlinear system--environment coupling selects an operator
space that is invisible to this direct environmental channel, rather
than a unique canonical generator. Different choices related by the
internal Bogoliubov reparameterization in Eq.~\eqref{eq:nonlinear_HSE_dark_reparametrization}
describe the same generated operator space. When the intrinsic system
Hamiltonian also satisfies the dynamical-invariance condition, this
operator space defines the corresponding dark mode.

For the remainder of this section we choose the passive representative
$D\equiv d$ and define the complementary environment-coupled mode
\begin{equation}
B=\frac{ca_{1}+a_{2}}{\sqrt{1+|c|^{2}}}.\label{eq:nonlinear_HSE_bright_mode}
\end{equation}
Using Eq.~\eqref{eq:nonlinear_HSE_conditions}, the nonlinear system--environment
coupling becomes 
\begin{equation}
H^{(2)}_{SE}=\chi_{2}(1+|c|^{2})B^{2}e^{\dagger}+\mathrm{H.c.}\label{eq:nonlinear_HSE_B2}
\end{equation}
Thus the auxiliary mode couples only to the collective pair operator
$B^{2}$ built from the environment-coupled mode $B$. The coupling
in Eq.~\eqref{eq:nonlinear_HSE_B2} contains no $D$ operator, so
every operator in the space generated by $D$ and $D^{\dagger}$ is
absent from the direct nonlinear system--environment coupling.

\subsection{Dynamical invariance under nonlinear system Hamiltonians}

After identifying the environment-invisible mode, we now ask which
intrinsic system Hamiltonians can preserve this mode without converting
it into the environment-coupled mode. We keep the mode operators 
\begin{equation}
B=\frac{ca_{1}+a_{2}}{\sqrt{1+|c|^{2}}},\qquad D=\frac{a_{1}-c^{*}a_{2}}{\sqrt{1+|c|^{2}}}.\label{eq:nonlinear_BD_modes}
\end{equation}
The system Hamiltonian is written as 
\begin{equation}
H_{S}=H^{(2)}_{S}+V_{\mathrm{nl}},\label{eq:nonlinear_full_HS}
\end{equation}
where the quadratic part has the form 
\begin{equation}
H^{(2)}_{S}=\omega_{1}a^{\dagger}_{1}a_{1}+\omega_{2}a^{\dagger}_{2}a_{2}+\kappa\left(a^{\dagger}_{1}a_{2}+a^{\dagger}_{2}a_{1}\right).\label{eq:nonlinear_quadratic_HS}
\end{equation}
To focus on the role of the nonlinear interactions, we assume that
this quadratic part already does not mix $D$ with $B$. As in Sec.~\ref{sec:passive_quadratic},
this requires 
\begin{equation}
2c\Delta+\kappa(1-c^{2})=0,\qquad\Delta=\frac{\omega_{1}-\omega_{2}}{2}.\label{eq:nonlinear_quadratic_decoupling}
\end{equation}
Equation~\eqref{eq:nonlinear_HSE_conditions} makes the nonlinear
system--environment coupling depend only on the collective pair operator
$B^{2}$, while Eq.~\eqref{eq:nonlinear_quadratic_decoupling} removes
quadratic $B$--$D$ conversion. The remaining issue is purely nonlinear:
does $V_{\mathrm{nl}}$ introduce processes that convert environment-invisible
excitations into environment-coupled ones?

\subsubsection{Self-Kerr, cross-Kerr, and pair-exchange interactions}

We first consider a standard quartic interaction family, 
\begin{equation}
\begin{aligned}V_{\mathrm{nl}} & =\frac{U}{2}\left(a^{\dagger2}_{1}a^{2}_{1}+a^{\dagger2}_{2}a^{2}_{2}\right)+V\,a^{\dagger}_{1}a_{1}a^{\dagger}_{2}a_{2}\\
 & \quad+\frac{P}{2}\left(a^{\dagger2}_{1}a^{2}_{2}+a^{\dagger2}_{2}a^{2}_{1}\right).
\end{aligned}
\label{eq:nl_natural_family}
\end{equation}
We take $U$, $V$, and $P$ to be real so that $V_{\mathrm{nl}}$
is Hermitian as written. The three terms represent local self-Kerr
nonlinearities, cross-Kerr interaction, and coherent pair exchange.
This is a natural starting point because these processes are among
the most common number-conserving quartic interactions in nonlinear
bosonic platforms. They also provide a minimal test of whether the
environment-invisible mode identified by the nonlinear channel remains
dynamically separated from the environment-coupled mode under the
intrinsic nonlinear dynamics.

Expressed in terms of the $B$ and $D$ mode operators, Eq.~\eqref{eq:nl_natural_family}
takes the form 
\begin{equation}
\begin{aligned}V_{\mathrm{nl}} & =\Lambda_{B}B^{\dagger2}B^{2}+\Lambda_{D}D^{\dagger2}D^{2}+\Lambda_{BD}B^{\dagger}BD^{\dagger}D\\
 & \quad+\left(\lambda_{2}B^{\dagger2}D^{2}+\mathrm{H.c.}\right)+\left(\lambda_{B1}B^{\dagger2}BD+\mathrm{H.c.}\right)\\
 & \quad+\left(\lambda_{D1}B^{\dagger}D^{\dagger}D^{2}+\mathrm{H.c.}\right),
\end{aligned}
\label{eq:nl_general_structure}
\end{equation}
with coefficients given in Appendix~\ref{app:nonlinear_two_mode}.
This decomposition displays the physical processes explicitly. The
first two terms are Kerr nonlinearities of the $B$ and $D$ modes
separately, where $B$ is the environment-coupled mode and $D$ is
the environment-invisible mode. The term proportional to $\lambda_{2}$
converts two $D$-mode excitations into two $B$-mode excitations,
and conversely. The terms proportional to $\lambda_{B1}$ and $\lambda_{D1}$
describe occupation-assisted conversion between the two modes. The
cross-Kerr term proportional to $\Lambda_{BD}$ does not exchange
excitations, but it makes the frequency of the $D$ mode depend on
the occupation of the $B$ mode. Since $B$ is coupled to the environmental
channel, such a dependence would transfer fluctuations of the environment-coupled
mode into the dynamics of the environment-invisible mode. For the
operators generated by $D$ to remain dynamically invariant without
involving $B$, this term must also be absent.

Thus for the $D$ mode to define a dark mode, all mixed $B$--$D$
terms must vanish: 
\begin{equation}
\Lambda_{BD}=\lambda_{2}=\lambda_{B1}=\lambda_{D1}=0.\label{eq:nl_no_mixed_terms}
\end{equation}
Solving these conditions within the quartic family of Eq.~\eqref{eq:nl_natural_family}
gives the nontrivial solution 
\begin{equation}
c=1,\qquad V=2U,\qquad P=U.\label{eq:nl_symmetric_manifold}
\end{equation}
Thus, for this standard set of quartic processes, dynamical invariance
is possible at the symmetric coupling point.

At $c=1$, the $B$ and $D$ modes are 
\begin{equation}
B=\frac{a_{1}+a_{2}}{\sqrt{2}},\qquad D=\frac{a_{1}-a_{2}}{\sqrt{2}},\label{eq:symmetric_BD_modes}
\end{equation}
and Eq.~\eqref{eq:nonlinear_quadratic_decoupling} reduces to $\Delta=0$.
Under Eq.~\eqref{eq:nl_symmetric_manifold}, the nonlinear interaction
becomes 
\begin{equation}
V_{\mathrm{nl}}=U\left(B^{\dagger2}B^{2}+D^{\dagger2}D^{2}\right).\label{eq:nl_symmetric_interaction_BD}
\end{equation}
The full system Hamiltonian is then 
\begin{equation}
\begin{aligned}H_{S} & =\omega_{B}B^{\dagger}B+\omega_{D}D^{\dagger}D\\
 & \quad+U\left(B^{\dagger2}B^{2}+D^{\dagger2}D^{2}\right).
\end{aligned}
\label{eq:nl_symmetric_full}
\end{equation}
The environment-invisible mode therefore survives genuine nonlinear
self-dynamics, provided the quartic interactions are balanced so that
no $B$--$D$ mixing remains.

\subsubsection{Asymmetric coupling and additional nonlinear pathways}

The symmetric case above is special. For a general value of $c$,
the self-Kerr, cross-Kerr, and pair-exchange terms in Eq.~\eqref{eq:nl_natural_family}
do not provide enough independent amplitudes to cancel all mixed terms
in Eq.~\eqref{eq:nl_general_structure}. Physically, the nonlinear
$B$--$D$ conversion pathways are then unbalanced. To restore exact
separation, one must introduce further nonlinear pathways whose amplitudes
can interfere destructively with the remaining mixed terms.

One convenient enlarged interaction is 
\begin{equation}
\begin{aligned}V^{(c)}_{\mathrm{nl}} & =U\Big[\alpha_{c}\left(a^{\dagger2}_{1}a^{2}_{1}+a^{\dagger2}_{2}a^{2}_{2}\right)+\beta_{c}\,a^{\dagger}_{1}a_{1}a^{\dagger}_{2}a_{2}\\
 & \qquad+\left(\gamma_{c}\,a^{\dagger2}_{1}a^{2}_{2}+\mathrm{H.c.}\right)\\
 & \qquad+\eta_{c}\left(c^{\ast}a^{\dagger2}_{1}a_{1}a_{2}-c\,a^{\dagger2}_{2}a_{1}a_{2}+\mathrm{H.c.}\right)\Big],
\end{aligned}
\label{eq:nl_extended_family}
\end{equation}
with 
\begin{equation}
\begin{aligned}\alpha_{c} & =\frac{1+|c|^{4}}{(1+|c|^{2})^{2}}, & \beta_{c} & =\frac{8|c|^{2}}{(1+|c|^{2})^{2}},\\
\gamma_{c} & =\frac{2(c^{\ast})^{2}}{(1+|c|^{2})^{2}}, & \eta_{c} & =\frac{2(|c|^{2}-1)}{(1+|c|^{2})^{2}}.
\end{aligned}
\label{eq:nl_extended_coeffs}
\end{equation}
The terms proportional to $\alpha_{c}$, $\beta_{c}$, and $\gamma_{c}$
generalize the self-Kerr, cross-Kerr, and pair-exchange processes
already present in Eq.~\eqref{eq:nl_natural_family}. The term proportional
to $\eta_{c}$ introduces correlated four-wave-mixing pathways. These
pathways vanish at the symmetric point $c=1$, but for asymmetric
collective couplings they supply the extra interference amplitudes
needed to cancel the remaining $B$--$D$ conversion terms.

Expressed in terms of the $B$ and $D$ mode operators, Eq.~\eqref{eq:nl_extended_family}
becomes 
\begin{equation}
V^{(c)}_{\mathrm{nl}}=U\left(B^{\dagger2}B^{2}+D^{\dagger2}D^{2}\right).\label{eq:nl_extended_family_BD}
\end{equation}
All mixed quartic terms cancel. Therefore, when the quadratic condition
Eq.~\eqref{eq:nonlinear_quadratic_decoupling} is also satisfied,
the full Hamiltonian again takes the separated form of Eq.~\eqref{eq:separated_Hamiltonian_form}.
The selected environment-invisible mode is then dynamically invariant
and therefore defines a dark mode.

These examples show that dark modes are not confined to linear models.
When the conversion pathways in a nonlinear environment coupling interfere
into a single collective two-photon process, the complementary mode
becomes invisible to the direct environment coupling. Once such a
mode has been identified, the condition on the system Hamiltonian
determines which intrinsic nonlinear interactions can maintain it,
and which additional processes are required when the simplest quartic
interactions are insufficient.

\section{Bogoliubov Dark Modes in Parametrically Driven Optomechanical Interfaces}

\label{sec:bogoliubov}

The nonlinear constructions in Sec.~\ref{sec:nonlinear_dark_modes}
show that dark modes can persist beyond the conventional quadratic-linear
framework. We now turn to another well-established physical realization
of the Hamiltonian conditions: the Bogoliubov dark mode in a parametrically
driven optomechanical interface \citep{poyatosQuantumReservoirEngineering1996,wangReservoirEngineeredEntanglementOptomechanical2013,tianRobustPhotonEntanglement2013,kronwaldArbitrarilyLargeSteadystate2013,woolleyTwomodeBackactionevadingMeasurements2013,woolleyTwomodeSqueezedStates2014}.
Unlike the passive collective modes discussed above, this dark mode
cannot be obtained as an annihilation-only linear combination within
the passive class. It emerges only after a Bogoliubov transformation
that mixes creation and annihilation operators.

We consider a single-cavity multimode optomechanical system supporting
two electromagnetic cavity modes, both coupled to the same mechanical
mode $m$ with frequency $\omega_{m}$. Mode 1 is driven near its
red mechanical sideband, while mode 2 is driven near its blue mechanical
sideband. After displacement by the strong classical drive fields,
linearization around the driven amplitudes, transformation to the
drive-rotating frame, and the rotating-wave approximation in the sideband-resolved
regime, the resulting resonant three-mode Hamiltonian can be written
as

\begin{equation}
H^{{\rm om}}_{{\rm lin}}=H^{{\rm om}}_{S}+H^{{\rm om}}_{E}+H^{{\rm om}}_{SE}.\label{eq:om_linearized_total}
\end{equation}
Here $a_{1}$ and $a_{2}$ denote the cavity fluctuation modes in
the drive frame. The effective cavity, mechanical, and interaction
contributions are 
\begin{equation}
H^{{\rm om}}_{S}=\omega_{m}\left(a^{\dagger}_{1}a_{1}-a^{\dagger}_{2}a_{2}\right),\label{eq:om_system_hamiltonian}
\end{equation}
\begin{equation}
H^{{\rm om}}_{E}=\omega_{m}m^{\dagger}m,\label{eq:om_environment_hamiltonian}
\end{equation}
and 
\begin{equation}
H^{{\rm om}}_{SE}=G_{1}\left(m^{\dagger}a_{1}+a^{\dagger}_{1}m\right)+G_{2}\left(ma_{2}+a^{\dagger}_{2}m^{\dagger}\right).\label{eq:om_system_environment}
\end{equation}
The couplings $G_{1}$ and $G_{2}$ are the drive-enhanced linearized
optomechanical couplings; their phases have been absorbed into the
definitions of the fluctuation operators. In this reduced description,
the two cavity fluctuation modes are the modes of interest, while
the mechanical mode serves as an auxiliary channel for the cavity
fluctuations. Its coupling to an external mechanical reservoir is
not written explicitly, since the dark-mode condition concerns the
coherent cavity--mechanical coupling in Eq.~\eqref{eq:om_system_environment}.
The first term in Eq.~\eqref{eq:om_system_environment} is the red-sideband
beam-splitter process, and the second is the blue-sideband two-mode-squeezing
process.

\begin{figure}[t]
\centering \includegraphics[width=0.95\linewidth]{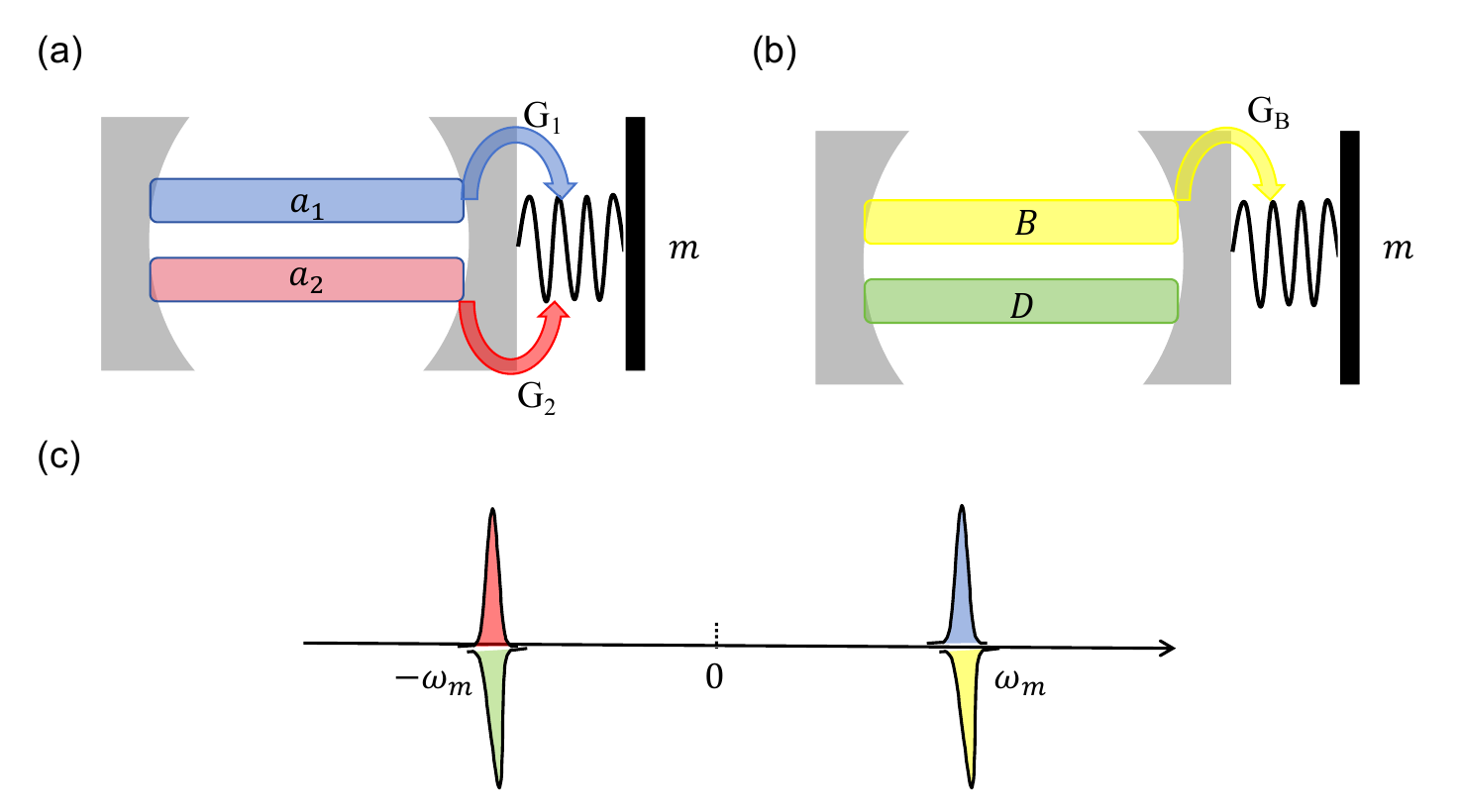} \caption{Bogoliubov dark mode in a parametrically driven optomechanical interface.
(a) In the effective drive-frame description, the cavity fluctuation
modes $a_{1}$ and $a_{2}$ couple to the same mechanical mode $m$
through the red-sideband beam-splitter coupling $G_{1}$ and the blue-sideband
two-mode-squeezing coupling $G_{2}$ in Eq.~\eqref{eq:om_system_environment}.
(b) In the Bogoliubov $B$--$D$ basis, only the bright mode $B$
couples to the mechanical channel, with effective strength $G_{B}=\sqrt{G^{2}_{1}-G^{2}_{2}}$,
while the mode $D$ is absent from this direct coupling. (c) Schematic
drive-frame frequency arrangement associated with the effective cavity
Hamiltonian in Eq.~\eqref{eq:om_system_hamiltonian}: the branch
associated with $a_{2}$ and $D$ lies near $-\omega_{m}$, while
the branch associated with $a_{1}$ and $B$ lies near $+\omega_{m}$.}
\label{fig:bogoliubov_dark_mode}
\end{figure}

The effective drive-frame coupling structure, the Bogoliubov transformation,
and the frequency arrangement are summarized schematically in Fig.~\ref{fig:bogoliubov_dark_mode}.

To search for a cavity mode that is invisible to the direct mechanical
coupling, we use the general two-mode Bogoliubov form 
\begin{equation}
D=u_{1}a_{1}+u_{2}a_{2}+w_{1}a^{\dagger}_{1}+w_{2}a^{\dagger}_{2}.\label{eq:om_general_dark_ansatz}
\end{equation}
The condition that this mode be absent from the direct mechanical
coupling is 
\begin{equation}
\left[H^{{\rm om}}_{SE},D\right]=\left[H^{{\rm om}}_{SE},D^{\dagger}\right]=0.\label{eq:om_dark_condition}
\end{equation}
Solving Eq.~\eqref{eq:om_dark_condition} within the Bogoliubov form
in Eq.~\eqref{eq:om_general_dark_ansatz}, and selecting the normalized
annihilation branch relevant for the stable regime $G_{1}>G_{2}$,
gives the convenient representative 
\begin{equation}
D=\frac{G_{2}a^{\dagger}_{1}+G_{1}a_{2}}{\sqrt{G^{2}_{1}-G^{2}_{2}}}.\label{eq:om_dark_mode}
\end{equation}
The derivation of Eq.~\eqref{eq:om_dark_mode}, together with the
normalization of the environment-invisible Bogoliubov solution, is
summarized in Appendix~\ref{app:om_dark_solution}. Introducing 
\begin{equation}
G_{B}=\sqrt{G^{2}_{1}-G^{2}_{2}},\qquad\tanh r=\frac{G_{2}}{G_{1}},\label{eq:om_GB_r}
\end{equation}
the environment-invisible mode can be written as 
\begin{equation}
D=a^{\dagger}_{1}\sinh r+a_{2}\cosh r.\label{eq:om_dark_mode_r}
\end{equation}
The complementary mode is 
\begin{equation}
B=\frac{G_{1}a_{1}+G_{2}a^{\dagger}_{2}}{\sqrt{G^{2}_{1}-G^{2}_{2}}}=a_{1}\cosh r+a^{\dagger}_{2}\sinh r.\label{eq:om_bright_mode}
\end{equation}
These modes satisfy 
\begin{equation}
{}[B,B^{\dagger}]=1,\qquad[D,D^{\dagger}]=1,\qquad[B,D]=[B,D^{\dagger}]=0.\label{eq:om_canonical_relations}
\end{equation}
Thus $B$ and $D$ form a pair of independent Bogoliubov modes.

The transformation from the cavity fluctuation modes to the Bogoliubov
modes is implemented by the two-mode squeezing unitary 
\begin{equation}
U_{r}=\exp\!\left[r\left(a^{\dagger}_{1}a^{\dagger}_{2}-a_{1}a_{2}\right)\right].\label{eq:om_squeezing_unitary}
\end{equation}
Indeed, 
\begin{equation}
B=U_{r}a_{1}U^{\dagger}_{r},\qquad D=U_{r}a_{2}U^{\dagger}_{r}.\label{eq:om_bogoliubov_transformation}
\end{equation}
This explicitly realizes the type of active canonical transformation
discussed in Sec.~\ref{sec:hamiltonian_conditions}. The relevant
modes are therefore not passive rotations of the cavity fields, but
Bogoliubov modes selected by the parametrically driven interaction.
Here the dark mode is not a passive superposition of annihilation
operators, but an active canonical mode involving both annihilation
and creation operators. Preparing or addressing such a mode requires
a phase-sensitive mode transformation, such as two-mode squeezing
generated by a nonlinear crystal, a Josephson parametric element,
or an equivalent driven bosonic interface. In this sense, a system
whose bare modes do not contain an obvious passive dark mode may still
support an engineered dark mode after an appropriate canonical transformation.

In the Bogoliubov basis, the system--environment interaction becomes
\begin{equation}
H^{{\rm om}}_{SE}=G_{B}\left(B^{\dagger}m+Bm^{\dagger}\right).\label{eq:om_bright_environment}
\end{equation}
The mechanical channel therefore couples only to the Bogoliubov mode
$B$. The mode $D$ is absent from the direct coupling to the mechanical
auxiliary mode.

This environment-invisible mode is not available within the passive
linear ansatz used in Sec.~\ref{sec:passive_quadratic}. If one restricts
the search to 
\begin{equation}
D_{p}=u_{1}a_{1}+u_{2}a_{2},\label{eq:om_passive_ansatz}
\end{equation}
the direct-decoupling condition has only the trivial solution 
\begin{equation}
u_{1}=u_{2}=0,\label{eq:om_passive_trivial}
\end{equation}
whenever $G_{1},G_{2}\neq0$. The environment-invisible cavity mode
therefore exists only in the active Bogoliubov class. Physically,
the red-sideband beam-splitter process and the blue-sideband two-mode-squeezing
process combine so that the mechanical mode couples to one collective
Bogoliubov cavity mode, while the complementary mode is invisible
to the mechanical mode.

It remains to verify that the system Hamiltonian does not convert
the mode $D$ into the environment-coupled mode $B$. In the Bogoliubov
basis, Eq.~\eqref{eq:om_system_hamiltonian} becomes 
\begin{equation}
H^{{\rm om}}_{S}=\omega_{m}\left(B^{\dagger}B-D^{\dagger}D\right).\label{eq:om_system_bogoliubov}
\end{equation}
Thus the system Hamiltonian contains no $B$--$D$ coupling. Equivalently,
\begin{equation}
\left[H^{{\rm om}}_{S},D\right]=\omega_{m}D,\qquad\left[H^{{\rm om}}_{S},D^{\dagger}\right]=-\omega_{m}D^{\dagger}.\label{eq:om_preservation_commutator}
\end{equation}
The commutators are proportional to the same mode and its adjoint.
Thus the intrinsic cavity dynamics does not rotate $D$ into the environment-coupled
mode $B$. Therefore, mode $D$ is indeed a dark mode.

This optomechanical example highlights a mechanism that is qualitatively
different from the passive constructions discussed above. In the passive
examples, the annihilation operator of a dark mode is formed only
from linear combinations of the annihilation operators of the bare
bosonic modes. Here, by contrast, the dark mode in this example necessarily
mixes creation and annihilation operators and is generated by a two-mode
squeezing transformation. This example therefore shows how the same
Hamiltonian conditions apply beyond passive linear combinations and
describe dark modes generated by active canonical transformations
in driven quantum optical systems.

\section{Discussion and Conclusion}

\label{sec:conclusion}

In this work, we have formulated Hamiltonian conditions for identifying
dark modes in multimode bosonic systems. The central point is that
a mode may be absent from the direct system--environment coupling,
but this alone does not guarantee that it remains dark under the full
dynamics. The intrinsic system Hamiltonian may still mix this mode
with modes that are exposed to the environment. A bosonic dark mode
therefore requires two conditions: environment invisibility under
the direct coupling and dynamical invariance under the system Hamiltonian.

We formulate these requirements for bosonic dark modes satisfying
the canonical commutation relations. The protected degree of freedom
is described not by a particular choice of annihilation operators,
but by the generated operator space. This space contains the observables
and processes associated with these modes. The system--environment
coupling must act trivially on this operator space, while the system
Hamiltonian must keep it dynamically invariant. The resulting dark-mode
dynamics may be collective rather than decomposable into independent
single-mode components.

When the environment-coupled and environment-invisible modes are completed
into a full bosonic mode basis, the second condition has a clear Hamiltonian
meaning. The system Hamiltonian may generate nontrivial dynamics within
the environment-coupled modes and within the environment-invisible
modes separately, but it must not contain coupling terms between the
environment-invisible modes and the environment-coupled modes.

The linear model provides the simplest benchmark for this situation.
For a system with linear system--environment coupling, environment-invisible
passive modes can be identified from the null space of the coupling
matrix. For a quadratic system Hamiltonian, one must further require
the corresponding subspace of environment-invisible collective modes
to be invariant under the Hamiltonian matrix. This reproduces the
standard structure of linear dark-mode theory and related decoherence-free
constructions. In the minimal two-mode model, the dynamical-invariance
condition reduces to an explicit mode-decoupling relation, which has
the same interference structure as the Friedrich--Wintgen mechanism
for bound states in the continuum.

We further applied the two conditions to nonlinear system--environment
coupling and nonlinear intrinsic interactions. When the system--environment
coupling is mediated by two-photon conversion into an auxiliary mode,
destructive interference among different conversion pathways can reduce
the environmental channel to a single collective two-photon process.
The complementary bosonic mode is then invisible to this nonlinear
channel. The intrinsic quartic Hamiltonian must satisfy an additional
interference requirement: it must not reintroduce mixed processes
that convert the environment-invisible mode into the environment-coupled
mode. Within the standard self-Kerr, cross-Kerr, and pair-exchange
family, this selects a symmetric coupling point. For asymmetric collective
couplings, additional correlated four-wave-mixing pathways provide
the extra amplitudes needed to cancel the remaining mixed nonlinear
conversion terms.

Finally, the parametrically driven optomechanical interface illustrates
a non-passive route to a dark mode. In the effective drive-frame description,
the red-sideband beam-splitter process and the blue-sideband two-mode-squeezing
process combine so that the mechanical auxiliary mode couples only
to one Bogoliubov bright mode of the cavity fluctuations. The mechanically
dark cavity mode is not an annihilation-only collective combination
of the bare cavity fluctuation modes. It necessarily mixes creation
and annihilation operators and is generated by a two-mode-squeezing
transformation. The intrinsic cavity Hamiltonian keeps the operator
space generated by this mode dynamically invariant, so the optomechanical
interface realizes an independently evolving Bogoliubov dark mode.

Taken together, these results provide a Hamiltonian method for both
identifying and engineering dark modes in multimode bosonic systems.
Given a physical model, the conditions can be used to determine whether
dark modes exist and to derive the corresponding annihilation and
creation operators. Conversely, starting from a desired mode to be
protected, the same conditions can guide the design of the full Hamiltonian
structure: the system--environment coupling should make this mode
invisible to the environment, while the intrinsic system Hamiltonian
should keep the generated operator space dynamically invariant. In
this sense, the Hamiltonian conditions provide a direct design principle
for realizing dark modes, applicable not only to passive linear systems
but also to nonlinear and driven quantum optical platforms.

Several directions remain open. It would be useful to develop systematic,
physically motivated ansätze or numerical methods for finding dark
modes when the relevant mode transformation is not known in advance.
Another important direction is to quantify how the separation degrades
under parameter mismatch, residual conversion between environment-invisible
and environment-coupled modes, or imperfect nonlinear balance. These
questions are directly relevant for using dark-mode interference as
a design principle in quantum optical, optomechanical, superconducting-circuit,
and other multimode bosonic platforms.
\begin{acknowledgments}
This work is supported by National Natural Science Foundation of China
(Grant No. 12075323), the Natural Science Foundation of Guangdong
Province of China (Grant No. 2025A1515011440) and the Innovation Program
for Quantum Science and Technology (Grant No. 2021ZD0300702).
\end{acknowledgments}

\appendix

\section{Generator Freedom and Collective Dark-Mode Dynamics}

\label{app:generator_freedom}

This appendix clarifies two related points about the operator space
$\mathcal{S}_{D}$. First, the degree of freedom that remains dark
is the operator space itself, rather than a particular choice of canonical
generators, or equivalently a particular choice of annihilation and
creation operators. The same operator space may therefore admit several
equivalent generator representations. Second, dynamical invariance
of the full operator space does not necessarily imply that its dynamics
can be decomposed into independently evolving single-mode dark components.
The dark-mode dynamics may therefore be collective rather than decomposable
into independent single-mode components.

\subsection{Generator freedom of the operator space}

Let 
\begin{equation}
\mathcal{S}_{D}=\left\{ F(D_{1},D^{\dagger}_{1},\ldots,D_{n_{D}},D^{\dagger}_{n_{D}})\right\} \label{eq:appB_dark_operator_set}
\end{equation}
be the operator space generated by a set of bosonic modes $\{D_{k}\}^{n_{D}}_{k=1}$.
The choice of generators is not unique. Another set of bosonic modes
$\{D'_{k}\}^{n_{D}}_{k=1}$ describes the same dark degrees of freedom
if it generates the same operator space, 
\begin{equation}
\left\{ F(D'_{1},D^{\prime\dagger}_{1},\ldots,D'_{n_{D}},D^{\prime\dagger}_{n_{D}})\right\} =\mathcal{S}_{D}.\label{eq:appB_same_operator_set}
\end{equation}
Since the Hamiltonian conditions depend only on the operators contained
in $\mathcal{S}_{D}$, 
\begin{equation}
{}[H_{SE},O_{D}]=0,\qquad[H_{S},O_{D}]\in\mathcal{S}_{D},\qquad\forall O_{D}\in\mathcal{S}_{D},\label{eq:appB_conditions_on_operator_set}
\end{equation}
any two generator choices that produce the same $\mathcal{S}_{D}$
represent the same dark-mode structure.

A common source of this freedom is a canonical reparameterization
internal to $\mathcal{S}_{D}$. For example, one may define 
\begin{equation}
D'_{k}=\sum^{n_{D}}_{\ell=1}\left(\mathsf{U}_{k\ell}D_{\ell}+\mathsf{V}_{k\ell}D^{\dagger}_{\ell}\right),\qquad k=1,\ldots,n_{D},\label{eq:appB_internal_Bogoliubov}
\end{equation}
provided the transformed modes satisfy the bosonic commutation relations
\begin{equation}
{}[D'_{k},D^{\prime\dagger}_{l}]=\delta_{kl},\qquad[D'_{k},D'_{l}]=0.\label{eq:appB_internal_CCR}
\end{equation}
When $\mathsf{V}=0$, this is an internal unitary rotation of the
generators. When $\mathsf{V}\neq0$, it is an internal Bogoliubov
reparameterization. Such changes do not alter the generated operator
space.

The single-mode freedom appearing in Sec.~\ref{sec:nonlinear_dark_modes}
is the simplest example. There, the nonlinear environment coupling
first selects a single operator space that is invisible to the direct
nonlinear system--environment coupling. Once the dynamical-invariance
condition is also imposed, this space becomes the corresponding dark
operator space. Its generator can be written as 
\begin{equation}
D=\mu d+\nu d^{\dagger},\qquad|\mu|^{2}-|\nu|^{2}=1.\label{eq:appB_single_mode_reparametrization}
\end{equation}
The modes $d$ and $D$ generate the same single-mode operator space.
They are therefore different canonical descriptions of the same dark
degree of freedom, not distinct physical dark modes.

\subsection{Collective dynamics versus independent dark modes}

It is important to distinguish dynamical invariance of the full operator
space from independent evolution of each single mode. The Hamiltonian
condition requires 
\begin{equation}
{}[H_{S},O_{D}]\in\mathcal{S}_{D},\qquad\forall O_{D}\in\mathcal{S}_{D}.\label{eq:appB_full_preservation}
\end{equation}
Equivalently, in a complete $D$--$B$ mode frame, the system Hamiltonian
may take the separated form 
\begin{equation}
\begin{split}H_{S}={} & H_{D}(D_{1},D^{\dagger}_{1},\ldots,D_{n_{D}},D^{\dagger}_{n_{D}})\\
 & +H_{B}(B_{1},B^{\dagger}_{1},\ldots,B_{n_{B}},B^{\dagger}_{n_{B}}).
\end{split}
\label{eq:appB_separated_form}
\end{equation}
This condition excludes conversion between the $D$ modes and the
environment-coupled $B$ modes. It does not, however, require the
dynamics inside the $D$ modes to split into independent one-mode
motions.

To isolate a single independently evolving dark mode, one would need
a mode $d\in\mathcal{S}_{D}$ whose own generated operator space 
\begin{equation}
\mathcal{S}_{d}=\left\{ F(d,d^{\dagger})\right\} \label{eq:appB_single_mode_operator_set}
\end{equation}
is dynamically invariant by itself, 
\begin{equation}
{}[H_{S},O_{d}]\in\mathcal{S}_{d},\qquad\forall O_{d}\in\mathcal{S}_{d}.\label{eq:appB_single_mode_preservation}
\end{equation}
This is a stronger requirement than Eq.~\eqref{eq:appB_full_preservation}.
The full operator space $\mathcal{S}_{D}$ may remain dark even when
no individual one-mode operator space $\mathcal{S}_{d}$ is invariant
on its own. In that case, the dark degree of freedom is a collective
multimode structure rather than a collection of independently evolving
modes.

A useful simplification occurs in special models with additional structure.
For example, in the passive quadratic limit of Sec.~\ref{sec:passive_quadratic},
the action of $H_{S}$ on passive dark modes is governed by the Hermitian
matrix $\Omega$ restricted to the environment-invisible coefficient
space. This restricted quadratic problem can be diagonalized by an
internal unitary rotation, yielding independently evolving dark normal
modes. This diagonalization is a special feature of the quadratic
setting and need not persist for more general nonlinear dark-mode
Hamiltonians.

\section{Generated Operator Space and the Absence of $D$--$B$ Mixing}

\label{app:closure}

This appendix justifies the equivalence between dynamical invariance
and the absence of mixed $D$--$B$ terms in the system Hamiltonian.
We use the same terminology as in the main text: $\mathcal{S}_{D}$
denotes the operator space generated by the dark modes, 
\begin{equation}
\mathcal{S}_{D}=\left\{ F(D_{1},D^{\dagger}_{1},\ldots,D_{n_{D}},D^{\dagger}_{n_{D}})\right\} .\label{eq:app_dark_operator_set}
\end{equation}
This space contains all observables and processes constructed only
from the $D$ modes.

After the $D$ modes are completed by complementary modes $\{B_{\alpha}\}^{n_{B}}_{\alpha=1}$
to a full bosonic mode basis, the condition of dynamical invariance
reads 
\begin{equation}
{}[H_{S},O_{D}]\in\mathcal{S}_{D},\qquad\forall O_{D}\in\mathcal{S}_{D}.\label{eq:app_dark_preservation}
\end{equation}
We show that this condition is equivalent, up to an additive scalar,
to a Hamiltonian with no mixed $D$--$B$ terms, 
\begin{equation}
\begin{split}H_{S}={} & H_{D}(D_{1},D^{\dagger}_{1},\ldots,D_{n_{D}},D^{\dagger}_{n_{D}})\\
 & +H_{B}(B_{1},B^{\dagger}_{1},\ldots,B_{n_{B}},B^{\dagger}_{n_{B}}).
\end{split}
\label{eq:app_separated_Hamiltonian}
\end{equation}
Here $H_{D}$ contains only $D$-mode operators, while $H_{B}$ contains
only $B$-mode operators.

The implication from Eq.~\eqref{eq:app_separated_Hamiltonian} to
Eq.~\eqref{eq:app_dark_preservation} is immediate. Since $H_{B}$
contains only $B$-mode operators, it commutes with every operator
in $\mathcal{S}_{D}$. Therefore, for any $O_{D}\in\mathcal{S}_{D}$,
\begin{equation}
{}[H_{S},O_{D}]=[H_{D},O_{D}]\in\mathcal{S}_{D}.\label{eq:app_forward_implication}
\end{equation}
Thus a Hamiltonian of the form in Eq.~\eqref{eq:app_separated_Hamiltonian}
cannot convert observables generated by the $D$ modes into operators
involving the $B$ modes.

We now prove the converse. In the complete $D$--$B$ mode representation,
any system Hamiltonian can be expanded in terms of linearly independent
$B$-mode operator structures. We write 
\begin{equation}
\begin{split}H_{S}=\sum_{\mu} & X_{\mu}(D_{1},D^{\dagger}_{1},\ldots,D_{n_{D}},D^{\dagger}_{n_{D}})\\
 & \times Y_{\mu}(B_{1},B^{\dagger}_{1},\ldots,B_{n_{B}},B^{\dagger}_{n_{B}}),
\end{split}
\label{eq:app_HS_DB_expansion}
\end{equation}
where the operators $Y_{\mu}$ are chosen to be linearly independent
$B$-mode operator structures, including the identity operator. The
factors $X_{\mu}$ are the corresponding $D$-mode operator functions.

For any $O_{D}\in\mathcal{S}_{D}$, all $Y_{\mu}$ commute with $O_{D}$,
and therefore 
\begin{equation}
{}[H_{S},O_{D}]=\sum_{\mu}[X_{\mu},O_{D}]\,Y_{\mu}.\label{eq:app_commutator_DB_expansion}
\end{equation}
Dynamical invariance requires this commutator to contain no nontrivial
$B$-mode operator for every $O_{D}\in\mathcal{S}_{D}$. Since the
$Y_{\mu}$ are linearly independent, every term with a nontrivial
$B$-mode factor must satisfy 
\begin{equation}
{}[X_{\mu},O_{D}]=0,\qquad\forall O_{D}\in\mathcal{S}_{D}.\label{eq:app_commutant_condition}
\end{equation}
An operator constructed from the $D$ modes that commutes with all
operators in $\mathcal{S}_{D}$ can only be a scalar. Therefore, for
every term with a nontrivial $B$-mode factor, 
\begin{equation}
X_{\mu}=c_{\mu}.\label{eq:app_X_scalar}
\end{equation}
Such terms are therefore $B$-mode terms only.

The remaining terms are those for which $Y_{\mu}$ is the identity.
They contain only $D$-mode operators and combine into $H_{D}$. The
terms with nontrivial $Y_{\mu}$ combine into $H_{B}$. Hence the
Hamiltonian can be written as Eq.~\eqref{eq:app_separated_Hamiltonian},
up to an additive scalar.

This proves the equivalence between dynamical invariance of the generated
operator space $\mathcal{S}_{D}$ and the absence of mixed $D$--$B$
terms in the system Hamiltonian. Physically, any Hamiltonian term
containing both a nontrivial $D$-mode operator and a nontrivial $B$-mode
operator would make the commutator of some $D$-mode observable depend
on the $B$ modes. Such a term is therefore incompatible with the
$D$-mode operators remaining dynamically invariant on their own.

\section{Coefficients of the Restricted Quartic Two-Mode Interaction}

\label{app:nonlinear_two_mode}

This appendix gives the explicit expansion of the restricted quartic
interaction family used in Sec.~\ref{sec:nonlinear_dark_modes} in
terms of the environment-coupled mode $B$ and the environment-invisible
mode $D$. The calculation supports the no-mixing conditions quoted
in the main text and shows that, within this restricted interaction
family, dynamical invariance of the $D$-mode operator space selects
the symmetric collective branch.

We consider the number-conserving quartic interaction 
\begin{equation}
\begin{aligned}V_{\mathrm{nl}} & =\frac{U}{2}\left(a^{\dagger2}_{1}a^{2}_{1}+a^{\dagger2}_{2}a^{2}_{2}\right)+V\,a^{\dagger}_{1}a_{1}a^{\dagger}_{2}a_{2}\\
 & \quad+\frac{P}{2}\left(a^{\dagger2}_{1}a^{2}_{2}+a^{\dagger2}_{2}a^{2}_{1}\right),
\end{aligned}
\label{eq:app_nl_restricted}
\end{equation}
where $U$, $V$, and $P$ are taken to be real. The three terms describe
self-Kerr nonlinearities, cross-Kerr interaction, and coherent pair
exchange. We use the collective modes 
\begin{equation}
B=\frac{ca_{1}+a_{2}}{\sqrt{1+|c|^{2}}},\qquad D=\frac{a_{1}-c^{\ast}a_{2}}{\sqrt{1+|c|^{2}}},\label{eq:app_BD_modes}
\end{equation}
where the nonlinear environment channel couples to the collective
mode $B$, while the complementary mode $D$ is absent from the direct
nonlinear environment coupling. Equivalently, 
\begin{equation}
a_{1}=\frac{c^{\ast}B+D}{\sqrt{1+|c|^{2}}},\qquad a_{2}=\frac{B-cD}{\sqrt{1+|c|^{2}}}.\label{eq:app_inverse_BD}
\end{equation}

Substituting Eq.~\eqref{eq:app_inverse_BD} into Eq.~\eqref{eq:app_nl_restricted}
and collecting normal-ordered terms gives 
\begin{equation}
\begin{aligned}V_{\mathrm{nl}} & =\Lambda_{B}B^{\dagger2}B^{2}+\Lambda_{D}D^{\dagger2}D^{2}+\Lambda_{BD}B^{\dagger}BD^{\dagger}D\\
 & \quad+\left(\lambda_{2}B^{\dagger2}D^{2}+\mathrm{H.c.}\right)+\left(\lambda_{B1}B^{\dagger2}BD+\mathrm{H.c.}\right)\\
 & \quad+\left(\lambda_{D1}B^{\dagger}D^{\dagger}D^{2}+\mathrm{H.c.}\right),
\end{aligned}
\label{eq:app_nl_restricted_BD}
\end{equation}
where 
\begin{equation}
\Lambda_{B}=\Lambda_{D}=\frac{U(1+|c|^{4})+2V|c|^{2}+P\left(c^{2}+(c^{\ast})^{2}\right)}{2(1+|c|^{2})^{2}},\label{eq:app_Lambda_BD_self}
\end{equation}
\begin{equation}
\Lambda_{BD}=\frac{V(1-|c|^{2})^{2}+4U|c|^{2}-2P\left(c^{2}+(c^{\ast})^{2}\right)}{(1+|c|^{2})^{2}},\label{eq:app_Lambda_BD_cross}
\end{equation}
\begin{equation}
\lambda_{2}=\frac{P(1+c^{4})+2c^{2}(U-V)}{2(1+|c|^{2})^{2}},\label{eq:app_lambda_2}
\end{equation}
\begin{equation}
\lambda_{B1}=\frac{P(c^{\ast}-c^{3})+(U-V)(c^{2}c^{\ast}-c)}{(1+|c|^{2})^{2}},\label{eq:app_lambda_B1}
\end{equation}
and 
\begin{equation}
\lambda_{D1}=\frac{P(c^{3}-c^{\ast})+(V-U)(c^{2}c^{\ast}-c)}{(1+|c|^{2})^{2}}.\label{eq:app_lambda_D1}
\end{equation}

For the environment-invisible mode $D$ to become a dark mode under
the intrinsic nonlinear dynamics, the operator space generated by
$D$, 
\begin{equation}
\mathcal{S}_{D}=\left\{ F(D,D^{\dagger})\right\} ,\label{eq:app_single_dark_operator_set}
\end{equation}
must remain dynamically invariant. In the expansion Eq.~\eqref{eq:app_nl_restricted_BD},
this requires all mixed $B$--$D$ terms to vanish: 
\begin{equation}
\Lambda_{BD}=\lambda_{2}=\lambda_{B1}=\lambda_{D1}=0.\label{eq:app_restricted_closure}
\end{equation}

Solving Eq.~\eqref{eq:app_restricted_closure} within the restricted
family Eq.~\eqref{eq:app_nl_restricted} gives the symmetric exact
branch 
\begin{equation}
c=1,\qquad P=U,\qquad V=2U.\label{eq:app_restricted_solution}
\end{equation}
Thus, within this restricted quartic model, the environment-invisible
mode selected by the nonlinear system--environment coupling is dynamically
invariant only on the symmetric collective branch quoted in Sec.~\ref{sec:nonlinear_dark_modes}.

\section{Direct Solution of the Optomechanical Bogoliubov Mode}

\label{app:om_dark_solution}

This appendix derives the Bogoliubov mode used in Sec.~\ref{sec:bogoliubov}
directly from the environment-invisibility condition. The calculation
shows how a cavity mode absent from the direct mechanical coupling
is obtained from a general Bogoliubov ansatz, and how the relative
size of $G_{1}$ and $G_{2}$ enters through the bosonic normalization.

We start from the linearized optomechanical coupling 
\begin{equation}
H^{{\rm om}}_{SE}=G_{1}\left(m^{\dagger}a_{1}+a^{\dagger}_{1}m\right)+G_{2}\left(ma_{2}+a^{\dagger}_{2}m^{\dagger}\right),\label{eq:app_om_HSE}
\end{equation}
where $G_{1},G_{2}>0$ have been chosen real by absorbing phases into
the definitions of the fluctuation operators. It is useful to write
Eq.~\eqref{eq:app_om_HSE} as 
\begin{equation}
H^{{\rm om}}_{SE}=m^{\dagger}L+mL^{\dagger},\qquad L=G_{1}a_{1}+G_{2}a^{\dagger}_{2}.\label{eq:app_om_L}
\end{equation}
Thus the mechanical mode couples directly to the Bogoliubov combination
$L$ of the two cavity modes.

We search for a cavity mode that is invisible to this direct mechanical
coupling using the general Bogoliubov ansatz 
\begin{equation}
D=u_{1}a_{1}+u_{2}a_{2}+w_{1}a^{\dagger}_{1}+w_{2}a^{\dagger}_{2}.\label{eq:app_om_general_ansatz}
\end{equation}
Since the mechanical mode commutes with all cavity operators, 
\begin{equation}
{}[H^{{\rm om}}_{SE},D]=m^{\dagger}[L,D]+m[L^{\dagger},D].\label{eq:app_om_commutator_HSE}
\end{equation}
The condition that $D$ be absent from the direct mechanical coupling
is therefore 
\begin{equation}
{}[L,D]=0,\qquad[L^{\dagger},D]=0.\label{eq:app_om_invisibility_conditions_L}
\end{equation}
Using 
\begin{equation}
L=G_{1}a_{1}+G_{2}a^{\dagger}_{2},\label{eq:app_om_L_explicit}
\end{equation}
we find 
\begin{equation}
{}[L,D]=G_{1}[a_{1},D]+G_{2}[a^{\dagger}_{2},D]=G_{1}w_{1}-G_{2}u_{2}.\label{eq:app_om_LD}
\end{equation}
Similarly, since 
\begin{equation}
L^{\dagger}=G_{1}a^{\dagger}_{1}+G_{2}a_{2},\label{eq:app_om_Ldagger}
\end{equation}
one obtains 
\begin{equation}
{}[L^{\dagger},D]=G_{1}[a^{\dagger}_{1},D]+G_{2}[a_{2},D]=-G_{1}u_{1}+G_{2}w_{2}.\label{eq:app_om_LdaggerD}
\end{equation}
The environment-invisibility conditions therefore give 
\begin{equation}
w_{1}=\frac{G_{2}}{G_{1}}u_{2},\qquad w_{2}=\frac{G_{1}}{G_{2}}u_{1}.\label{eq:app_om_coeff_relations}
\end{equation}
Substituting these relations into Eq.~\eqref{eq:app_om_general_ansatz},
the general cavity mode invisible to the direct mechanical coupling
can be written as 
\begin{equation}
D=u_{1}\left(a_{1}+\frac{G_{1}}{G_{2}}a^{\dagger}_{2}\right)+u_{2}\left(a_{2}+\frac{G_{2}}{G_{1}}a^{\dagger}_{1}\right).\label{eq:app_om_general_solution}
\end{equation}

The remaining step is to impose the bosonic normalization. For the
operator in Eq.~\eqref{eq:app_om_general_ansatz}, 
\begin{equation}
{}[D,D^{\dagger}]=|u_{1}|^{2}+|u_{2}|^{2}-|w_{1}|^{2}-|w_{2}|^{2}.\label{eq:app_om_norm_general}
\end{equation}
Using Eq.~\eqref{eq:app_om_coeff_relations}, this becomes 
\begin{equation}
{}[D,D^{\dagger}]=\left(1-\frac{G^{2}_{1}}{G^{2}_{2}}\right)|u_{1}|^{2}+\left(1-\frac{G^{2}_{2}}{G^{2}_{1}}\right)|u_{2}|^{2}.\label{eq:app_om_norm_constraint}
\end{equation}

In the stable parameter regime used in the main text, 
\begin{equation}
G_{1}>G_{2},\label{eq:app_om_stable_regime}
\end{equation}
the coefficient of $|u_{2}|^{2}$ in Eq.~\eqref{eq:app_om_norm_constraint}
is positive, while the coefficient of $|u_{1}|^{2}$ is negative.
A positive-norm annihilation operator is therefore obtained most simply
by setting $u_{1}=0$ and keeping the $u_{2}$ branch: 
\begin{equation}
D\propto a_{2}+\frac{G_{2}}{G_{1}}a^{\dagger}_{1}.\label{eq:app_om_positive_branch}
\end{equation}
With 
\begin{equation}
G_{B}=\sqrt{G^{2}_{1}-G^{2}_{2}},\label{eq:app_om_GB}
\end{equation}
the normalized environment-invisible Bogoliubov mode is 
\begin{equation}
D_{0}=\frac{G_{1}a_{2}+G_{2}a^{\dagger}_{1}}{G_{B}}.\label{eq:app_om_dark_mode}
\end{equation}
Equivalently, introducing 
\begin{equation}
\tanh r=\frac{G_{2}}{G_{1}},\qquad\cosh r=\frac{G_{1}}{G_{B}},\qquad\sinh r=\frac{G_{2}}{G_{B}},\label{eq:app_om_r_definitions}
\end{equation}
one obtains 
\begin{equation}
D_{0}=a^{\dagger}_{1}\sinh r+a_{2}\cosh r,\label{eq:app_om_dark_mode_r}
\end{equation}
which is the Bogoliubov mode used in Eq.~\eqref{eq:om_dark_mode}.
Together with the dynamical-invariance condition verified in the main
text, this mode defines the optomechanical Bogoliubov dark mode.

Equation~\eqref{eq:app_om_general_solution} also shows that the
normalized representative in Eq.~\eqref{eq:app_om_dark_mode} is
not the only possible choice of generator for the same single-mode
operator space. For $G_{1}>G_{2}$, a general canonical generator
may be written as 
\begin{equation}
D'=\mu D_{0}+\nu D^{\dagger}_{0},\qquad|\mu|^{2}-|\nu|^{2}=1.\label{eq:app_om_generator_freedom}
\end{equation}
In terms of the coefficients in Eq.~\eqref{eq:app_om_general_solution},
this corresponds to 
\begin{equation}
\mu=\frac{G_{B}}{G_{1}}u_{2},\qquad\nu=\frac{G_{B}}{G_{2}}u_{1}.\label{eq:app_om_mu_nu}
\end{equation}
Thus choices with both $u_{1}$ and $u_{2}$ nonzero are allowed whenever
Eq.~\eqref{eq:app_om_norm_constraint} is normalized to unity. These
choices generate the same single-mode operator space as $D_{0}$,
in agreement with the generator freedom discussed in Appendix~\ref{app:generator_freedom}.

At the boundary 
\begin{equation}
G_{1}=G_{2},\label{eq:app_om_boundary}
\end{equation}
Eq.~\eqref{eq:app_om_norm_constraint} gives 
\begin{equation}
{}[D,D^{\dagger}]=0\label{eq:app_om_zero_norm}
\end{equation}
for every operator of the form Eq.~\eqref{eq:app_om_general_solution}
that is invisible to the direct mechanical coupling. Hence, although
such operators commute with the mechanical coupling, they cannot be
normalized into bosonic annihilation operators. In this sense, the
Bogoliubov dark mode used in the main text ceases to exist at $G_{1}=G_{2}$.

If instead $G_{2}>G_{1}$, the positive-norm branch is obtained by
exchanging the roles of the two terms in Eq.~\eqref{eq:app_om_norm_constraint}.
A canonical representative can then be chosen as 
\begin{equation}
\widetilde{D}=\frac{G_{2}a_{1}+G_{1}a^{\dagger}_{2}}{\sqrt{G^{2}_{2}-G^{2}_{1}}}.\label{eq:app_om_opposite_branch}
\end{equation}
This is the conjugate branch relative to the one used in the main
text. It is included here for completeness; the stable optomechanical
interface considered in Sec.~\ref{sec:bogoliubov} is the branch
$G_{1}>G_{2}$, for which Eq.~\eqref{eq:app_om_dark_mode} provides
the normalized environment-invisible Bogoliubov mode.

\bibliography{bosonic}

\end{document}